\documentstyle[12pt,epsf]{article}
\def\singlespace 
{\smallskipamount=3.75pt plus1pt minus1pt
\medskipamount=7.5pt plus2pt minus2pt
\bigskipamount=15pa plus4pa minus4pt \normalbaselineskip=12pt plus0pt
minus0pt \normallineskip=1pt \normallineskiplimit=0pt \jot=3.75pt
{\def\smallskip {\vskip\smallskipamount}} {\def\medskip
{\vskip\medskipamount}} {\def\bigskip {\vskip\bigskipamount}}
{\setbox\strutbox=\hbox{\vrule height10.5pt depth4.5pt width 0pt}}
\parskip 7.5pt \normalbaselines} 

\def\middlespace
{\smallskipamount=5.625pt plus1.5pt minus1.5pt \medskipamount=11.25pt
plus3pt minus3pt \bigskipamount=22.5pt plus6pt minus6pt
\normalbaselineskip=22.5pt plus0pt minus0pt \normallineskip=1pt
\normallineskiplimit=0pt \jot=5.625pt {\def\smallskip
{\vskip\smallskipamount}} {\def\medskip {\vskip\medskipamount}}
{\def\bigskip {\vskip\bigskipamount}} {\setbox\strutbox=\hbox{\vrule
height15.75pt depth6.75pt width 0pt}} \parskip 11.25pt
\normalbaselines} 

\def\doublespace 
{\smallskipamount=7.5pt plus2pt minus2pt \medskipamount=15pt plus4pt
minus4pt \bigskipamount=30pt plus8pt minus8pt \normalbaselineskip=30pt
plus0pt minus0pt
\normallineskip=2pt \normallineskiplimit=0pt \jot=7.5pt
{\def\smallskip {\vskip\smallskipamount}} {\def\medskip
{\vskip\medskipamount}} {\def\bigskip {\vskip\bigskipamount}}
{\setbox\strutbox=\hbox{\vrule height21.0pt depth9.0pt width 0pt}}
\parskip 15.0pt \normalbaselines}

\newcommand{\be}{\begin{equation}}
\newcommand{\ee}{\end{equation}}
\newcommand{\bea}{\begin{eqnarray}}
\newcommand{\eea}{\end{eqnarray}}
      
 \def\a{\alpha} \def\b{\beta}

\begin{document}
\middlespace

\vskip 2cm
\begin{flushright} SINP/TNP/01-21 \\
\underbar{Accepted in Phys. Rev. D.}
\end{flushright}
\begin{center}
\Large {\bf Adjoint bulk scalars and supersymmetric
unification in the presence of extra dimensions} \\ 
\vskip 1cm Biswajoy Brahmachari
\footnote{{\tt electronic address:biswajoy@theory.saha.ernet.in}}\\
\end{center}
\begin{center}
Theoretical Physics Group\\
Saha Institute of Nuclear Physics\\
1/AF Bidhannagar, Kolkata-700064, India
\\

\end{center}
\vskip 2cm 
{
%\singlespace
\begin{center}
\underbar{Abstract} \\
\end{center}

There are several advantages of introducing adjoint superfields
at intermediate energies around $M=10^{13}$ GeV. Such as (i) 
gauge couplings still unify (ii) neutrino masses and mixings are
produced (iii) primordial lepton asymmetry can be produced. We point out that 
if adjoint scalars have bulk excitations along with gauge bosons whereas 
fermions and the doublet scalar live on boundary then $N=2$ supersymmetric 
beta functions $\tilde{b_i}$ vanish. Thus even if extra dimensions 
open up at an intermediate scale $\mu_0$ and all $N=2$ Yang-Mills fields 
as well as $N=2$ matter fields in the adjoint representation propagate
in the bulk, still gauge couplings renormalize beyond $\mu_0$ just 
like they do in 4-dimensions with adjoint scalars. Consequently
unification is achieved in the presence to extra dimensions, mass scales
are determined uniquely via Renormalization Group Equations(RGE) and 
unification scale remains high enough to suppress proton decay.
This scenario can be falsified if we get signatures of extra
dimensions at low energy.

\newpage

We know that the gauge group of Glashow-Weinburg-Salam(GSW) standard model
is $G \equiv SU(3)_c \times SU(2)_L \times U(1)_Y$. Now suppose after
compactifications string theory
gives us a group $G \times G$ and the scalars are only in fundamental
representation of each component group. Then during the symmetry breaking
$G \times G \rightarrow G^{diag}$ to the diagonal $G$ we obtain scalars
in the adjoint representation. For example consider the symmetry
breaking $SU(3)\times SU(3) \longrightarrow SU(3)^{diag}$. Nine scalars 
of (3,3) representation transform after symmetry breaking as 1 + 8 of 
$SU(3)^{diag}$. The singlet can get a VEV for the symmetry to break and 
the adjoint octet is produced. Furthermore (1,3) or (3,1) representation 
become fundamental representation of $SU(3)^{diag}$.

It was noted by Bachas Fabre and Yanagida\cite{bfy} that when we add
the matter superfields (8,1,0)+(1,3,0)+(1,1,0) to Minimal
Supersymmetric Standard Model(MSSM) spectrum
at a scale of around $10^{12-13}$ GeVs, the gauge couplings still unify
but the unification scale is pushed up from $10^{16}$ GeVs to the string
scale of around $10^{18}$ GeVs. This case becomes more interesting
from the current experimental results on the neutrino masses as the
fermionic partners of (1,3,0) and (1,1,0) scalars are capable of
producing tree level diagrams giving Majorana masses to neutrinos\cite{maprl}.
This mass is of the order of $m^2_Z/M$ where M is the mass of the
adjoint fermions. Thus to get the mass of neutrino in the 1 eV range
the scale $M$ is required to be $10^{13}$ GeV as suggested by the
Renormalization Group(RG) analyses\cite{ma}. Note that the triplet T and the 
singlet S has the gauge invariant Yukawa couplings $L H_2 S$ and 
$L H_2 T$. Following this line it can also been shown that out of 
equilibrium decay of either S or T could produce a tiny lepton asymmetry 
in the early universe which leads to leptogenesis\cite{ma}.

Suppose $\delta$ numbers of extra space-time dimensions open up at a 
scale $\mu_0$ which is much below the Planck scale
of $10^{19}$ GeVs. In this case the running of the gauge couplings
beyond $\mu_0$ no longer remains logerithmic. Furthermore beyond
$\mu_0$ the Yang-Mills and matter superfields are arranged 
in $N=2$ multiplets. Thus $N=1$ beta functions $b_i$ for the 
gauge couplings are to be supplemented by $N=2$ beta functions 
$\tilde{b}_i$. Thus $a~priori$ it is not at all guaranteed that 
the gauge couplings will preserve unification. In this context a 
very important work of Dines Dudas and Gherghetta(DDG)\cite{ddg} shows 
that for MSSM the gauge 
couplings do not only unify in 4-dimensions which is well-known but also
unify in the presence of extra dimensions present at low energy
\cite{antoniadis}. This unification
is independent of the scale $\mu_0$ and the number of
extra dimensions $\delta$. Indeed this is a remarkable property of 
the low energy particle content of MSSM. Following this paper
a number of recent studies in this direction have been
performed\cite{carone,quiros,andreja,more}. Beyond the
gauge structure of MSSM grand unification with intermediate
left-right symmetry has also been considered\cite{lorenzana}. 

Following this line we wish to ask what are the properties of MSSM enhanced
by adjoint matter fields ? Does it preserve unification of gauge 
couplings in the presence of enlarged extra dimensions? 
Furthermore after unification of couplings can we uniquely predict
right mass scale of extra states such that correct neutrino mass
can be generated ? Clearly for us a new scale has been introduced which is
$\mu_0$. In this paper we will analyze this case and find out
a specific embedding of the superfields in higher dimensional
space-time which will give rise to gauge coupling unification. Note 
that there exists a non-trivial change in the nature of gauge
coupling evolution beyond $\mu_0$ as the adjoint scalars remain in 
the bulk and hence the
$SU(3)_c \times SU(2)_L \times U(1)_Y$ gauge bosons also
live in the bulk. This is due to the string constraint that bulk 
fields transform under bulk gauge groups; this issue has been discussed 
by Carone\cite{carone} for example.  

An $N=2$ theory consists of $N=2$ Yang-Mills fields as
well as $N=2$ matter fields. $N=2$ Yang-Mills theory
has $N=1$ Yang-Mills fields plus one Wess-Zumino
multiplet in the adjoint representation of
the Yang-Mills gauge group. $N=2$ matter consists of
Wess-Zumino multiplets in $R_i$ and $\overline{R_i}$
representations. Thus for each additional adjoint matter representation
below $\mu_0$ that we are going to introduce we will have their
Kaluza-Klein (KK) excitations leading effectively to a pair 
of adjoints above $\mu_0$ to complete the $N=2$ hypermultiplet.
This is because adjoint representations are self-conjugate.
This is in the same spirit as of fermions where members of each
pair above $\mu_0$ has opposite charges. Thus vector-like fermion 
pairs above $\mu_0$ not only complete
$N=2$ hypermultiplet but also cancel anomaly. The $N=2$ 
beta function coefficients are expressed as\cite{west}
\be
\tilde{b}_i= 2[-C_2(G_i) + \sum_{i} T_i]. \label{n2beta}
\ee
Here $C_2(G_i)$ is the quadratic casimir of the i th group $G_i$.
If $G_i$ is SU(N) then $C_2(G_i)=N$. 
There are $\eta$ number of fermion generations experiencing extra 
dimensions and they contribute via the second term. Then we can 
explicitly isolate fermion contributions and write
\be
\tilde{b}_i= 2[-C_2(G_i) + \sum_{S} T_S] + 4 \eta. \label{n2b}
\ee
Now the sum exclude fermion generations. To check DDG case let us combine
doublet Higgs into a $N=2$ multiplet (in this case it contains two low 
energy N=1 Higgs doublets\footnote{Rather than combining two Higgs
fields into an $N=2$ multiplet another possibility
would be to augment each Higgs field to its own $N=2$ multiplets.}) 
in the bulk along with SU(2) gauge bosons. 
We immediately get $\tilde{b_2}=-3+ 4 \eta$ as in Reference\cite{ddg}. 
At this point suppose the adjoint scalars live in the bulk whereas the 
doublet scalar and fermions live on the boundary then we get\footnote{
To cross check, see Table[2] of Reference\cite{andreja}: 
$\Delta \tilde{b_2}$=4 for (1,3,0) and $\Delta \tilde{b_3}$=6 for (8,1,0)}
\be
\tilde{b}_i= 0. \label{zero}
\ee
Now we must examine the running of the gauge couplings beyond $\mu_0$
scale.
\be
\a^{-1}_i(\Lambda)= \a^{-1}_i(\mu_0)-{b_i \over 2 \pi} 
ln[{\Lambda \over \mu_0}]
       + {\tilde{b_i} \over 2 \pi} ln[{\Lambda \over \mu_0}]
           - {\tilde{b_i} X_\delta \over 2 \pi \delta}
         [({\Lambda \over \mu_0 })^\delta -1]. \label{running1}
\ee
Here $b_i$ are $N=1$ coefficients including the adjoint scalars.
$\Lambda$ is the scale at which the couplings are being evaluated,
$\delta$ is the number of extra dimensions, $X_\delta$ is given by, 
\be
X_\delta= { 2 \pi^{\delta/2} \over \delta\Gamma(\delta/2)},
\ee
and, $\Gamma$ is the Euler gamma function.
Thus we have implicitly assumed that $\mu_0 > M$, where M is the
mass scale from which the adjoint scalars start contributing to
RG analysis. We have
\be
b_1=33/5,~~b_2=3,~~b_3=0.
\ee
Using Eqn. (\ref{zero}) the running equation between
$\mu_0$ and $\Lambda$ in Eqn. (\ref{running1}) reduces to,
\be
\a^{-1}_i(\Lambda)= \a^{-1}_i(\mu_0)-{b_i \over 2 \pi} 
ln[{\Lambda \over \mu_0}].
\label{running2}
\ee
Now it is trivial to supplement Eqn. (\ref{running2}) with the running of the
gauge couplings from $m_Z$ to $M$ using the MSSM beta functions 
$\beta_i^{MSSM}$ and from $M$ to $\mu_0$ using $b_i$. Finally we obtain,
\be
\a^{-1}_i(\Lambda)= \a^{-1}_i(m_Z)
-{\beta^{MSSM}_i \over 2 \pi} ln[{M \over m_Z}]
-{b_i \over 2 \pi} ln[{\mu_0 \over M}]
-{b_i \over 2 \pi} ln[{\Lambda \over \mu_0}].
\label{running3}
\ee
This equation can be simplified to
\be
\a^{-1}_i(\Lambda)= \a^{-1}_i(m_Z)
-{\beta^{MSSM}_i \over 2 \pi} ln[{M \over m_Z}]
-{b_i \over 2 \pi} ln[{\Lambda \over M}].
\label{running4}
\ee
Thus as long as $\mu_0 > M$ the scale $\mu_0$ dissapears from the
renormalization group analysis. The coefficients $\beta^{MSSM}_i$
are given by\cite{jones},
\be
\b^{MSSM}_1=33/5,~~\b^{MSSM}_2=1,~~\b^{MSSM}_3=-3. \label{bmssm}
\ee 
Using the low energy experimental values of the three couplings
it is very easy to solve equations (\ref{running4}) for unification scale 
$M_X$, intermediate scale $M$ and unification coupling $\alpha_X$.
We obtain one-loop result $\alpha^{-1}_X =20.45, M_X = 10^{18}$ GeV 
and $M = 10^{12.85}$ GeV which matches well with\cite{mar}. However
above $\mu_0$ we have $N=2$ supermultiplets thus it is sufficient to
carry out one-loop RG analysis above $\mu_0$. Because in this 
case $\mu_0 > M$ we obtain, 
\be
\mu_0 > 10^{12.85} {\rm ~~GeV}. \label{cond} 
\ee
Note as long as the condition (\ref{cond}) is satisfied, independent
of the value of $\mu_0$ and independent of the number of extra 
dimensions the unification scale $M_X=10^{18}$ remains invariant.
Similarly independent of the scale $\mu_0$ the intermediate
scale is fixed by RGE to be $M=10^{12.85}$. These are the
main results of this paper. If we had power low running
we would not have achieved a splitting of $10^5$ GeVs
between $\mu_0$ and $M_X$.  

Now the phenomenological motivation of this paper will be clear.
It was shown recently \cite{ma} that a fermion triplet and a fermion
singlet at around $10^{13}$ GeV gives neutrino masses and mixings
that explain solar and atmospheric neutrino oscillations.
Furthermore the out of equilibrium decay of the triplet
to $H_2$ and L gives rise to a lepton asymmetry of the
universe. We see that these nice features can be maintained
in a similar unified model in the presence of extra dimensions.

Next we ask what if $\mu_0 << M$ ? In this case we can consider
bulk excitation of the doublet Higgses so
using Eqn. (\ref{n2beta}) we get $N=2$ supersymmetric
beta functions, 
\be
\tilde{b_1}=3/5 + 4 \eta,~~\tilde{b_2}=-3 + 4 \eta ,
~~\tilde{b_3}=-6 + 4 \eta,
\ee
where $\eta$ is the number of generations experiencing extra dimensions.
Thus along with the MSSM coefficients we recover the scenario
proposed by Dines Dudas and Gherghetta\cite{ddg,quiros} which gives unification
close to the scale $\mu_0$. Variations of this scenario with minimal
particle content can also be found\cite{carone}. There also exist 
detailed study of unification scale by introducing various extra 
multiplets\cite{carone,andreja}. 
Note that in Table 13 of Reference \cite{andreja} representations
contained upto 75 of SU(5) are discussed. These representations
actually include the adjoint fields of zero hypercharge
that we are considering. However one
extra state at a time was included in the RG analysis and
thus highest unification scale was found for the introduction
of (3,6,1/3) fermion at $M_X=1.3 \times 10^{14}$ GeV.
We can also entertain the possibility that below
the scale $\mu_0$ two copies of (8,1,0)+(1,3,0)+(1,1,0)
survives. Then $N=1$ beta functions are
\be
b_1=33/5,~~b_2=5,~~b_3=3.
\ee
In this case we get $M=10^{15.42}$, $M_X=10^{18}$ and $\alpha^{-1}_X=
20.45$. Note that one of the main reasons on introducing (1,3,0)+(1,1,0)
is to produce neutrino mass\cite{maprl} via see-saw
mechanism\cite{seesaw}. In the case of $M=10^{15.42}$
GeV we get that neutrinos can be at most as heavy as $10^{-2}$ eV.
Thus it will not be possible to address LSND results\cite{lsnd} where
$\Delta m_{e\mu}^2 \sim 1 eV^2$ is required. They may
also be too light for the purpose of atmospheric
neutrino oscillation\cite{atm} where $\Delta m^2\sim 10^{-2}$ is
the best fit region. Furthermore
for neutrinos to become the hot component of dark matter\cite{hdm}
its mass needs to be in the 1 eV range. 

Is it possible to have $\mu_0 < M < M_X$ ? We 
have not found any such scenario with the choice of our fields that 
is MSSM supplemented by adjoint scalars. If more extra particles
are added it may be possible to produce delicately balanced $\tilde{b}$
and $b$ coefficients to produce $\mu_0 < M < M_X$. This possibility
is beyond the scope of present work because we are adding only
adjoint superfields which are natural light remnants of string 
compactifications.

On proton decay one can take one of  two viewpoints.
Interesting ideas have been forwarded where compactifications
is done on a $Z_2$ orbifold where interactions that lead to
proton decay vanish at orbifold fixed points. For 
example we can assign a discrete reflection symmetry depending on
the extra dimensions under which the propagators mediating
proton decay are odd whereas all fermions are even\cite{ddg}.
A more extended discrete symmetry depending on extra space-time dimensions
has been also suggested recently in
Reference\cite{alta} where matter fields have $Z_2 \times Z^\prime_2$
charge assignments which forbids proton decay. One can also
consider gauging and breaking an additional U(1) symmetry on a distant
brane\cite{brane}. These are higher dimensional solutions
of proton decay problem. Otherwise in the worse case if these 
mechanisms are absent one must have the traditional solution of 
large $M_X > 10^{16}$ GeV and get propagator suppression. In the 
second case the results obtained in this paper will be quite useful
in the presence of enlarged extra dimensions with $\mu_0 > 10^{12.85}$ GeV.

In summary, radiative corrections in supersymmetric theories in the
presence of extra dimensions come from two sources. Fields remaining
on the boundary as well as fields propagating in the 
bulk give logerithmic corrections. These corrections are
are of the form ${(b -\tilde{b}) \over 2 \pi} ln({\Lambda \over \mu_0})$
where $b$ is $N=1$ beta functions. Fields propagating in the bulk 
form $N=2$ multiplets. 
We note that power law corrections to gauge couplings come exclusively
from the fields propagating to the bulk. In the case where
we have adjoint hypermultiplets propagating in the bulk
whereas fermions as well as doublet Higgses staying on the boundary,
$N=2$ beta functions vanish making the $N=2$ part finite.
Now as power law corrections to gauge couplings are proportional
to $N=2$ beta functions $\tilde{b}$ they are
absent. Thus in our example radiative corrections to gauge couplings maintain
their logerithmic nature even in the presence of enlarged extra dimensions
that is when $\Lambda > \mu_0$ as displayed in Equation (\ref{running3}).
Unification is delayed till $M_X=10^{18}$ GeV. This our main 
result which is independent of number of extra dimensions and the 
scale $\mu_0$ as long as $\mu_0 > 10^{12.85}$ GeV. 

A theory can become $N=2$ finite in other ways too. For example
if we add 2N fundamental scalars to the theory, $\tilde{b}$
will also vanish. However logerithmic corrections imparted
by 2N fundamental representations of matter above $\mu_0$
will not give gauge coupling unification. This can be easily
understood by examining 4th term in the right hand
side of Equation (\ref{running3}). Hence our example is unique 
where $N=2$ beta functions vanish and also gauge couplings unify
in the presence of extra dimensions.

Another thing we must notice that when $\mu_0 << M $ we
have accelerated unification when doublet Higgs is
in the bulk. Whereas if $\mu_0 > M$ we have logerithmic
unification. Thus the nature of evolution changes
in a nontrivial way when $\mu_0$ crosses $M$. 
Thus Equation (\ref{cond}) which is a result of
RG analysis is a strong constraint
on our scenario and it can also be used to falsify our
scenario. We know that there are a number of studies
which focus on experimental signatures of 
large extra dimensions at the TeV scale\cite{signature}. If we
get positive signature of extra dimensions at low
energy our scenario will be ruled out.

Finally as we have conventional running, we get many
advantages of GUTs. The proton decay rate is suppressed
because of large $M_X$. The scale M plays the role
of the see-saw scale and we can get neutrino mass
in the 1 eV region as suggested by neutrino
oscillation experiments. Neutrino can also be
a dark matter candidate in this case. The adjoint
triplet and singlet can decay via Yukawa interactions
and can lead to Leptogenesis. It will be an 
interesting idea to examine RGE of Yukawa
couplings in this scenario and calculate
fermion mass hierarchies.  

Communications from R. N. Mohapatra and a number of
comments from M. Quiros are gratefully acknowledged.


\begin{thebibliography}{99}                                                

\bibitem{bfy}C. Bachas, C. Fabre, T. Yanagida, Phys. Lett. 
{\bf B370}, 49 (1996) .


\bibitem{maprl}R. Foot, H. Lew, X. G. He, G. C. Joshi, Z. Phys.
{\bf C44}, 441 (1989); E. Ma, Phys. Rev. Lett {\bf81}, 1171 (1998).


\bibitem{ma} B. Brahmachari, E. Ma, U. Sarkar, e-Print Archive: 
hep-ph/0105278, Phys. Lett. B. (to appear).


\bibitem{ddg}
  K. R. Dienes, E. Dudas, T. Gherghetta, Phys. Lett. {\bf B436}, 55 (1998);
 Nucl. Phys. {\bf B537}, 47 (1999).

\bibitem{antoniadis} I. Antoniadis, Phys. Lett. {\bf B246} 377 (1990);
I. Antoniadis, K. Benakli, Phys. Lett. {\bf B326}, 69 (1994);
I. Antoniadis, K. Benakli, M. Quiros, Phys.Lett. {\bf B331}, 313 (1994).


\bibitem{quiros} A. Delgado, M. Quiros, Nucl. Phys.
{\bf B559}, 235 (1999); Phys. Lett. {\bf B484}, 355 (2000)


\bibitem{carone}C. D. Carone, Phys. Lett. {\bf B454}, 70 (1999).


\bibitem{andreja} 
P. H. Frampton, A. Rasin, Phys. Lett. {\bf B460}, 313 (1999). 

\bibitem{more}
N. Borghini, Y. Gouverneur, M. H.G. Tytgat, e-Print Archive:
hep-ph/0108094; F. Feruglio, e-Print Archive: hep-ph/0105321;
J. Kubo, H. Terao, G. Zoupanos, e-Print Archive: hep-ph/0010069;
H. Cheng, B. A. Dobrescu, C. T. Hill, Nucl. Phys. {\bf B573}, 597 (2000);
K. Huitu, T. Kobayashi, Phys. Lett. {\bf B470}, 90 (1999);
M. Masip, Phys. Rev. {\bf D62}, 065011, (2000);
D. Dumitru, S. Nandi, Phys. Rev. {\bf D62}, 046006 (2000);
E.G. Floratos, G.K. Leontaris, Phys. Lett. {\bf B465}, 95 (1999);
Guy F. de Teramond, Phys. Rev. {\bf D60}, 095010 (1999);
T. Kobayashi, J. Kubo, M. Mondragon, G. Zoupanos,
Nucl. Phys. {\bf B550}, 99 (1999); 
S.A. Abel, S.F. King, Phys. Rev. {\bf D59}, 095010 (1999); D. Ghilencea, 
G.G. Ross, Phys. Lett. {\bf B442}, 165, (1998); Nucl. Phys. {\bf B569},
391 (2000).
 


\bibitem{lorenzana} A. Perez-Lorenzana, R.N. Mohapatra, 
Nucl. Phys.{\bf B559}, 255 (1999).


\bibitem{west} P. West, ``{\it Introduction to supersymmetry and
supergravity}'' (World Scientific, Singapore, 1990);
P.P. Srivastava, ``{\it Supersymmetry, superfields and 
supergravity: an introduction}'' (Hilger Bristol UK 1986)

\bibitem{jones} J.E. Bjorkman, D.R.T. Jones, 
Nucl. Phys. {\bf B259}, 533 (1985);
U. Amaldi, W. de Boer, H. Furstenau, Phys. Lett. {\bf B260}, 447 (1991). 

\bibitem{mar}
M. Bastero-Gil, B. Brahmachari, Phys. Lett. {\bf B403}, 51 (1997) 
T. Han, T. Yanagida, R.J. Zhang, Phys. Rev. {\bf D58}, 095011 (1998);  
J.L. Chkareuli, C.D. Froggatt, I.G. Gogoladze, A.B. Kobakhidze,
Nucl. Phys. {\bf B594}, 23 (2001).


\bibitem{seesaw}
M. Gell-Mann, P. Rammond and R. Slansky, in {\it Supergravity}, edited
by P. Van Nieuwenhuizen and D. Z. Freedman, (North-Holland,
Amsterdam, 1979), p.~315; T. Yanagida, in {\it Proceedings of the Workshop
on the UNified Theory and the Baryon Number in the Universe}, edited by
O. Sawada and A. Sugamoto (KEK, Tsukuba, Japan, 1979), p.~95; R. N. Mohapatra
and G. Senjanovi\'c. Phys. Rev. Lett {\bf 44}, 912 (1980).


\bibitem{alta}
G. Altarelli, F. Feruglio, Phys. Lett. {\bf B511}, 257 (2001). 

\bibitem{brane}
I. Antoniadis, N. Arkani-Hamed, S. Dimopoulos, G.R. Dvali, Phys. Lett.
{\bf B436}, 257 (1998); G. Shiu, S.H.H. Tye, Phys. Rev. {\bf D58}, 
106007 (1998); 
Z. Kakushadze,  Nucl. Phys. {\bf B548}, 205 (1998); Nucl. 
Phys. {\bf B552} 3 (1999). 

\bibitem{lsnd}
LSND Collaboration, C. Athanassopoulos {\it et. al.}, Phys. Rev. Lett.
{\bf 75}, 2650 (1995); {\it ibid.} 3082 (1996); {\it ibid.} 81 1774
(1998); Phys. Rev. {\bf C 58}, 2489 (1998).

\bibitem{atm} S. Fukuda {\it et al.}, Super-Kamiokande Collaboration, Phys.
Rev. Lett. {\bf 85}, 3999 (2000) and references therein.

\bibitem{hdm}
D. O. Caldwell, UCSB-HEP-99-18, in `Baltimore 1999, Neutrinos in the new
millennium' 331-340 e-Print Archive: hep-ph/9910349; 
GENIUS Collaboration, Published in `Marina del Rey 2000, Sources and
detection of dark matter and dark energy in the universe' 493-504 
e-Print Archive: astro-ph/0005568.  

\bibitem{signature}
   Bob Olivier, Invited talk at 36th Rencontres de
   Moriond on QCD and Hadronic Interactions, Les Arcs, France, 17-24 Mar
   2001, LPNHE 2001-04, e-Print archive: hep-ex/0108015;
   D.Bourilkov, Invited LEP2 review talk,
   Les Rencontres de Physique de la Vallee d'Aoste, La Thuile, Aosta
   Valley, Italy, 4-10 March 2001, e-Print archive: hep-ex/0103039;
   T. Ferbel, presented at Cairo International Conference on 
    High Energy Physics (CICHEP), 9 - 14 January 2001, Cairo, Egypt, 
    e-Print archive: hep-ex/0103009; C. D. Hoyle, U. Schmidt, 
    B. R. Heckel, E. G. Adelberger, J. H. Gundlach, D. J. Kapner, 
    H. E. Swanson, Phys. Rev. Lett. {\bf 86}, 1418,  (2001);
    D0 Collaboration, Phys. Rev. Lett. {\bf 86}, 1156 (2001);
    H1 Collaboration, C.Adloff, et al, Phys. Lett. 
    {\bf B479}, 358 (2000).
   

\end{thebibliography}
\end{document}